\documentclass[journal, 11pt, draftcls, onecolumn]{IEEEtran}

\usepackage{amsmath}
\usepackage{amssymb}
\usepackage{times}
\usepackage{color}
\usepackage{graphicx}
\usepackage{cite,url}
\usepackage{pst-all}

\newcommand{\insertfig}[4]{
\begin{figure}[ht]
\centerline{\includegraphics[width=#1\columnwidth]{#2.eps}}
\caption{#3}\label{#4}\end{figure}}

\DeclareMathAlphabet{\mathsfbf}{OT1}{cmss}{sbc}{n}

\newcommand{\example}[2]{
\begin{center}
\parbox{0.9\columnwidth}{
\rule{0.9\columnwidth}{0.5mm}\\
\noindent {\bf Example~#1:} 
#2\\
\rule{0.9\columnwidth}{0.5mm}
}
\end{center}
}

\newtheorem{lemma}{Lemma}[section]

\newcommand{\EE}{\mathop{\mathbb{E}}\limits} 
\newcommand{\RR}{\mathbb{R}} 
\newcommand{\ZZ}{\mathbb{Z}} 
\newcommand{\Herm}{^\dagger} 
\newcommand{\Tran}{^{\rm T}} 

\newcommand{\ee}{{\rm e}}
\newcommand{\jj}{{\rm j}}  
\newcommand{\dd}{{\rm\,d}} 

\newcommand{\av}{{\bf a}}

\newcommand{\nv}{{\bf n}}

\newcommand{\pv}{{\bf p}}
\newcommand{\qv}{{\bf q}}
\newcommand{\rv}{{\bf r}}
\newcommand{\sv}{{\bf s}}

\newcommand{\wv}{{\bf w}}

\newcommand{\xv}{{\bf x}}
\newcommand{\yv}{{\bf y}}
\newcommand{\zv}{{\bf z}}
\newcommand{\zerov}{{\bf 0}}

\newcommand{\ellv}{\boldsymbol{\ell}}
\newcommand{\Am}{{\bf A}}
\newcommand{\Bm}{{\bf B}}
\newcommand{\Cm}{{\bf C}}

\newcommand{\Gm}{{\bf G}}

\newcommand{\Id}{{\bf I}}

\newcommand{\Lm}{{\bf L}}

\newcommand{\Rm}{{\bf R}}
\newcommand{\Sm}{{\bf S}}
\newcommand{\Tm}{{\bf T}}
\newcommand{\Um}{{\bf U}}
\newcommand{\Wm}{{\bf W}}

\newcommand{\Xm}{{\bf X}}

\newcommand{\Hc}{{\cal H}}

\newcommand{\Lc}{{\cal L}}

\newcommand{\Pc}{{\cal P}}
\newcommand{\Qc}{{\cal Q}}

\newcommand{\Xc}{{\cal X}}

\newcommand{\omegav}{\boldsymbol{\omega}}

\newcommand{\Lambdam}{\hbox{\boldmath$\Lambda$}}

\newcommand{\sinc}{{\hbox{sinc}}}

\def\trace{\mathsf{Tr}}

\def\ben{\begin{enumerate}}
\def\beq{\begin{equation}}
\def\beqa{\begin{eqnarray}}
\def\bit{\begin{itemize}}
\def\een{\end{enumerate}}
\def\eeq{\end{equation}}
\def\eeqa{\end{eqnarray}}
\def\eit{\end{itemize}}

\def\non{\nonumber\\}

\def\union{\mathop{\cup}\limits}

\def\MSEinf{{\rm MSE}_{\infty}}

\def\limbeta{\lim_{\substack{M,r\rightarrow +\infty \\\beta}}}

\title{Reconstruction of Multidimensional Signals from Irregular Noisy Samples 
\thanks{This work was supported by MIUR through the 
MEADOW project}}

\author{Alessandro Nordio {\em Member IEEE}, Carla-Fabiana
  Chiasserini {\em Member IEEE}, Emanuele Viterbo {\em Member IEEE}
}

\begin{document}
\maketitle

\begin{abstract}
We focus on a multidimensional field 
with uncorrelated spectrum, and study the quality 
of the reconstructed signal when the field samples are 
irregularly spaced and affected by independent and identically
distributed noise.  
More specifically, we apply linear reconstruction techniques and 
take the mean square error (MSE) of the field estimate  as a metric to
evaluate the signal reconstruction quality.
We find that the MSE analysis could be carried out by using the
closed-form expression of the eigenvalue distribution of the matrix
representing the sampling system. Unfortunately, such distribution is
still unknown. Thus, we first derive a closed-form expression of the
distribution moments, and we find that  the eigenvalue distribution
tends to the Mar\v{c}enko-Pastur distribution as the field dimension
goes to infinity. Finally, by using our approach, we 
derive a tight approximation to the MSE of the reconstructed field.
\end{abstract}

\section{Introduction}
\label{sec:introduction}

We address the important issue of reconstructing a multidimensional signal  
from a collection of samples that are noisy and not uniformly spaced.
As a case study, we consider a wireless sensor network
for environmental monitoring, where the nodes sensing 
the physical phenomenon (hereinafter also called  
field) are randomly deployed over the area under observation.
The sensors sample a $d$-dimensional spatially finite physical field, where 
$d$ may take into account
spatial dimensions as well as the temporal dimension. 
Examples of such fields are pressure or temperature, on a 4-dimension domain, i.e., 
three spatial coordinates plus the time dimension.
A  spatially finite physical field is
not bandlimited, however it 
admits an infinite Fourier series expansion.
Here, we consider a finite approximation of the physical field obtained 
by truncating such series, assuming that
the contribution of the truncated terms is negligible.

In our case study, 
we assume that the measured samples are transferred from the sensors to a 
common data-collecting unit, the so-called sink node, which
is in charge of reconstructing the field. Distributed systems,
where in-network processing, is performed are out
of the scope of this work. 
We do not deal with issues related to 
information transport and, thus, we 
assume that all samples are correctly
received at the sink node. 
The field samples, however,  are corrupted
by additive i.i.d. noise, due to quantization, round-off errors or
quality of the sensing device.
Furthermore, the sampling points are known at 
the sink node, because {\em (i)} either sensors are located at 
pre-defined positions or their position can be 
estimated through a localization technique~\cite{Moore04}, and
{\em (ii)} the sampling time is either periodic  or included in the information 
sent to the sink.

Several efficient and fast algorithms have been proposed
to numerically reconstruct or approximate a signal in such setting, 
which amount to the solution of a linear system 
(see \cite{Feichtinger93,Feichtinger95} and references therein).
A widely used technique consists in processing the sensors' measures by means of 
a linear filter, which is a function of the system parameters known 
at the sink. 
We observe that the following two major factors affect the
linear reconstruction: 
\begin{itemize}
\item[{\em (i)}] the given machine precision, which may prevent the reconstruction 
algorithm from performing correctly  and may lead to a  non-negligible probability of
reconstruction failure \cite{Feichtinger95}, 
\item[{\em (ii)}] the noise level affecting the sensors' measurements.
\end{itemize}
In the latter case, a measure of reconstruction accuracy is given by the mean square error
(MSE) of the field estimate. 
In~\cite{Tyrrenian,infocom07}, 
we have found that these issues could be studied by 
using the eigenvalue distribution of the reconstruction matrix; however, 
obtaining such a distribution is still an open problem.
In this work, we first extend the system model and the problem formulation presented
in \cite{infocom07} to the case of multidimensional fields (Section \ref{sec:preliminaries}). 
Then, we derive a closed-form expression of the  
moments of the eigenvalue distribution, through asymptotic 
analysis (Section \ref{sec:moments}). By using the moments expressions, we prove that 
the eigenvalue distribution of the matrix representing the sampling system
tends to the Mar\v{c}enko-Pastur distribution \cite{MarcenkoPastur} 
as the field dimension $d \rightarrow \infty$
(Section \ref{sec:approx-distribution}).

We apply our results to the study of the MSE of the field estimate,
when the sensors measurements are noisy and the reconstruction at the
sink is performed through linear filtering.

We generalize the MSE expressions to the multidimensional case (with finite $d$), and we show 
that, by using the Mar\v{c}enko-Pastur distribution instead of the actual 
eigenvalue distribution, we obtain
an approximation to the MSE of the reconstructed field which is very tight 
for $d\ge 3$ (Section \ref{sec:exploitation}).

Before providing a detailed description of our analysis, 
in the next section we discuss some related studies and highlight our
main contributions with respect to previous work.

\section{Related work and main contributions\label{sec:related}}

Relevant to our work is the literature on
spectral analysis, where, however, 
several studies deal with
regularly sampled signals
(e.g.,~\cite{Stoica} and references therein).  
An excellent guide to irregular sampling 
is \cite{Marvasti}, which presents a large number of techniques, 
algorithms, and applications. 
Reconstruction techniques for irregularly or randomly
sampled signals can be found in
\cite{Feichtinger95,Rauhut1,FeichtingerGrochenig}, just to name few.
In particular, Feichtinger and Gr\"ochenig in~\cite{FeichtingerGrochenig} provide an
error analysis of an iterative reconstruction algorithm taking into account round-off errors, jitters, 
truncation errors and aliasing.
From the theoretical point of view, irregular sampling has been studied in 
\cite{Feichtinger95,Rauhut1,Rauhut2,CandesTao,FeichtingerGrochenig,AldroubiGrochenig,Grochenig99} 
and references therein.

In the context of sensor networks,
efficient techniques for spatial sampling are 
proposed in~\cite{Perillo04,Willett04}. 
In particular, 
in~\cite{Willett04}, an adaptive sampling is 
described, which allows the central data-collector to vary the 
number of active sensors, i.e., samples, according to the desired 
resolution level.
Data acquisition is also studied in~\cite{Kumar03}, 
where the authors consider a unidimensional field, uniformly sampled at 
the Nyquist frequency by low precision sensors. The authors show that 
the number of samples can be traded-off with the 
precision of sensors.
The problem of the reconstruction of a bandlimited signal
from an irregular set of samples at unknown locations is addressed
in \cite{Marziliano00}. There, different solution methods are proposed, and 
the conditions for which there exist multiple solutions or a unique solution 
are discussed. Differently from~\cite{Marziliano00}, we assume that
the sink can either acquire or estimate the sensor locations
and that sensors are randomly deployed. 

The field reconstruction at the sink node with spatial and
temporal correlation among sensor measures is studied, for instance, in 
\cite{CristescuVetterli,Poor,Ganesan,Vuran04,Rachlin1}.
Other interesting studies can be found in~\cite{Zhao,Early}, which address
the perturbations of regular sampling in shift-invariant spaces~\cite{Zhao} and
the reconstruction of irregularly sampled images in presence of measurement 
noise~\cite{Early}.

We point out that our main contribution with respect to previous 
work on signal sampling and reconstruction is the probabilistic approach we adopt 
to analyze  the quality level of a signal reconstructed from a set of
irregular, noisy samples.
Our analysis, however, applies to sampling systems where the field reconstruction
is performed in a centralized manner.
Finally, we highlight that 
our previous work~\cite{infocom07} assumes that 
sensors are uniformly distributed over 
the spatial observation interval and may be displaced around a known average
location.  The effects of
noisy measures and jittered positions are analyzed when linear reconstruction 
techniques are employed. However, only the unidimensional case is studied and 
semi-analytical derivations of the MSE of the reconstructed field are obtained.
In \cite{ipsn07}, instead, sensors are assumed to be fixed, and  the objective 
is to  evaluate the performance 
of a linear reconstruction technique in the presence of quasi-equally spaced
sensor layouts.

\subsection{Main results}

The goal of this work is to provide an analytical study
on the reconstruction quality of 
a multidimensional physical field,
with uncorrelated spectrum.
The field samples
are (i) irregularly spaced, since they are gathered by a 
randomly deployed sensor network  and
(ii) affected by i.i.d. noise.
The sink node receives the field samples and runs the 
reconstruction algorithm in a centralized manner.  
Our major contributions with respect to previous work are as follows.
\begin{itemize}
\item [1.] Given a $d$-dimensional problem formulation, we
obtain analytical expressions for the moments of the eigenvalue
distribution of the reconstruction matrix.
Using the expressions of the moments, we show that the eigenvalue
distribution tends to the Mar\v{c}enko-Pastur distribution \cite{MarcenkoPastur} 
as the field dimension 
$d \rightarrow \infty$.
\item[2.] We apply our results to the study of the quality of a reconstructed field 
and derive a tight approximation to the MSE of the estimated field. 

\end{itemize}


\section{Preliminaries\label{sec:preliminaries}} 

We first present the multidimensional formulation
of our reconstruction problem. Then, we give
some background on linear reconstruction techniques and
generalize  to the multidimensional case some results 
previously obtained in
the unidimensional case \cite{infocom07}.
Finally, we highlight the main steps followed in our 
study.\\

{\em Notation:} Lower case bold letters denote column vectors, while upper case 
bold letters denote matrices.
We denote the $(h,k)$-th entry of the matrix $\Xm$ by $(\Xm)_{h,k}$, 
the transpose of $\xv$ by $\xv^{\rm T}$, and the conjugate transpose of $\xv$
by $\xv^{\dagger}$.
The identity matrix is denoted by $\Id$.
Finally, $\EE[x]$ is the average of $x$ and subscripts to 
the average operator specify the variable with respect to which the average
is taken.

\subsection{Irregular sampling of multidimensional signals}

Let us consider a $d$-dimensional, spatially-finite  physical field
($d \geq 1$), where $r$ sensors are located in the hypercube $\Hc = \{
\xv\,|\,\xv \in [0,1)^d\}$ and measure the value of the field. 
We assume that the sensor sampling points are known. 
At first, we consider that they are deterministic, then we will assume
that they are i.i.d. random variables uniformly distributed in the
hypercube $\Hc$.  

When observed over a finite region, a $d$-dimensional physical field
$s(\xv)$ with finite energy $E_s$ admits an infinite $d$-dimensional
Fourier series expansion with coefficients $\tilde{a}_{\ellv}$, such
that $E_s = \sum_{\ell_1,\ldots,\ell_d=-\infty}^{+\infty} |\tilde{a}_{\ellv}|^2$
where $\ellv=[\ell_1,\ldots,\ell_d]$ is a vector of integers and
$\ell_m$, $m=1,\ldots,d$ represents the index of the expansion along
the $m$-th dimension. 
We truncate the expansion to $2M+1$ terms per dimension where $M$ is
such that 
$\sum_{\ell_1,\ldots,\ell_d=-\infty}^{-M-1}|\tilde{a}_{\ellv}|^2
  +\sum_{\ell_1,\ldots,\ell_d=M+1}^{+\infty}|\tilde{a}_{\ellv}|^2 \ll E_s$
Therefore, one can think of $M$ as the approximate one-sided bandwidth
(per dimension) of the field, which can be approximated over the
finite region $\Hc$ as  
  \begin{equation}
  \label{eq:blsig}
   s(\xv) = (2M+1)^{-d/2}\sum_{\ellv}
  a_{\nu(\ellv)} \ee^{\jj 2\pi  \xv^{\rm T} \ellv}
   \end{equation}
where the term $(2M+1)^{-d/2}$ is a normalization factor and
$\sum_{\ellv}$ represents a $d$-dimensional sum over the vector
$\ellv$, with $\ell_m =-M,\ldots, M$.
Also, $a_{\nu(\ellv)}=\tilde{a}_{\ellv}$ and the function
\[\nu(\ellv) = \sum_{m=1}^d (2M+1)^{m-1} \ell_m, \]
$-\frac{(2M+1)^d-1}{2} \le \nu(\ellv) \le +\frac{(2M+1)^d-1}{2}$ maps
the vector $\ellv$ onto a scalar index. Note that, while
$\tilde{a}_{\ellv}$ has a vectorial index, $a_{\nu(\ellv)}$ has a
scalar index and it has been introduced to simplify the notation.
As an example, for $d=2$ and $M=1$, we have $\nu(\ellv) =
3\ell_1+\ell_2$ and 
$s(x_1,x_2) = \frac{1}{3}\sum_{\ell_1=-1}^1\sum_{\ell_2=-1}^1
a_{3\ell_2+\ell_1}\ee^{ \jj 2\pi(x_1\ell_1+x_2\ell_2)}$.

Let $\Xc= \{\xv_1, \ldots, \xv_r\},$ with 
$\xv_q=[x_{q,1},\ldots,x_{q,d}]^{\rm T} \in \Hc$, $q=1, \ldots, r$,
be the set of sampling points, and $\sv=[s_1,\ldots,s_r]^{\rm T}$, 
$s_q=s(\xv_q)$, the values of the corresponding  field samples. 
Following~\cite{Feichtinger95}, we write the vector of field values
$\sv$ as a function of the spectrum: 
\begin{equation}
\label{eq:s}
\sv= \Gm_d^{\dagger}\av 
\end{equation}
where $\av$ is a vector of size $(2M+1)^d$, whose
  $\nu(\ellv)$-th entry is given by $a_{\nu(\ellv)}$,
and $\Gm_d$ is the $\left(2M+1\right)^d \times r$  matrix:
\begin{equation}
\label{eq:G}
(\Gm_d)_{\nu(\ellv),q}=(2M+1)^{-d/2}\ee^{-\jj 2\pi  \xv_q^{\rm T} \ellv}
\end{equation}  
In general, the entries of $\av$ can be correlated with covariance
matrix $\EE[\av \av\Herm]= \sigma^2_a\Cm_a$, and ${\rm
  Tr}\{\Cm_a\}=(2M+1)^d$. In the following, we restrict our analysis
to the class of fields characterized by $\EE[\av \av\Herm]=
\sigma^2_a\Id$. If the sensor measurements, $\pv = [p_1,\ldots,
  p_r]^{\rm T}$, are noisy, then the relation between sensors' samples
and field spectrum can be written as:  
\begin{equation}
\pv = \sv + \nv = \Gm_d^\dagger\av + \nv
\label{eq:p}
\end{equation}
where the noise is represented by the $r$-size, zero-mean random vector $\nv$,
with covariance matrix $\EE[\nv\nv\Herm] = \sigma^2_n\Id_{r}$.
We define the signal-to-noise ratio on the measure as:
${\rm SNR}_m  \overset{\triangle}{=} \sigma^2_a/\sigma^2_n 
\overset{\triangle}{=} 1/\alpha$.

\subsection{Sampling rate}
Following~\cite{infocom07}, we introduce the parameter $\beta$ defined as:
\begin{equation}
\label{eq:beta}
\beta= \frac{(2M+1)^d}{r}
\end{equation}
This parameter represents the ratio between the number of harmonics used
for the field reconstruction and  the number of sensors sampling the field.
In the following, we consider $\beta\in \left[0,1\right)$. 
Notice that for fixed $\beta$ and $M$, the number $r$ of samples
exponentially increases with $d$.
  
\subsection{Previous results on reconstruction quality}

Given an estimate $\hat{\av}$ of the field spectrum $\av$, the reconstructed signal is:
\begin{equation}
\label{eq:blsighat}
\hat{s}(\xv) = (2M+1)^{-d/2}\sum_{\ellv}  \hat{a}_{\nu(\ellv)} \ee^{\jj 2\pi  \xv^{\rm T} \ellv}
\end{equation}
As reconstruction performance metric, we consider the MSE of the field estimate, which, for
any given set of sampling points $\Xc$, is defined as: 
\begin{equation}
{\rm MSE}_\Xc 
=\EE_{\av, \nv}\int_{\Hc} \left|\hat{s}(\xv)-s(\xv)\right|^2\dd \xv
=\frac{\EE_{\av, \nv}\|\hat{\av}-\av\|^2}{(2M+1)^d}
\label{eq:mse}
\end{equation}
where the average is taken with respect to the
subscripted random vectors.  
Note that (\ref{eq:mse})  still assumes that the sampling points are deterministic; 
this assumption will be removed later in the paper.

For linear models such as~(\ref{eq:p}), several estimation techniques based on linear
filtering are available in the literature~\cite{LibroEstimation}.
We employ a filter
$\Bm$ such that the estimate of the field spectrum is given by the linear operation
\begin{equation}
\hat{\av} = \Bm\Herm\pv
\label{eq:hat_a}
\end{equation}
where $\Bm$ is an $r \times (2M+1)^d$ matrix.
In particular, we consider the linear filter providing the best performance in terms of MSE,
i.e., the linear minimum MSE (LMMSE) filter%
\footnote{Notice that when the covariance matrix of $\av$ is known, the 
LMMSE filter generalizes 
to $(\Gm_d\Herm\Cm_a\Gm_d  +\alpha \Id)^{-1} \Gm_d\Herm\Cm_a$.}
\cite{LibroEstimation}:
\begin{equation}
\Bm = \Gm_d \Herm(\Rm_d +\alpha \Id)^{-1}
\label{eq:B}
\end{equation}
where $\Rm_d= \Gm_d\Gm_d\Herm$.

From now on, we carry out our analysis under the assumption that
the elements of the set $\Xc$ are independent random vectors,
with i.i.d. entries, uniformly distributed in the hypercube $\Hc$.

In~\cite{infocom07}, we have shown that a simple and effective tool to 
evaluate the performance of large finite systems
is asymptotic analysis. We computed the MSE by letting the field number of harmonics  and the
number of samples  grow to infinity, while their ratio $\beta=(2M+1)^d/r$ is kept
constant. We observed the validity of asymptotic analysis results, even for
small values of $M$ and $r$.
Similarly, here we consider as performance metric the
{\em asymptotic average} MSE, normalized to $\sigma^2_a$:
\begin{equation}
\MSEinf = \limbeta\frac{1}{\sigma^2_a}\EE_\Xc[{\rm MSE}_\Xc] 
\label{eq:MSEinf-persample}
\end{equation}
where $\beta$ below the limit denotes the ratio which is kept constant. 
In (\ref{eq:MSEinf-persample}), 
the average is over all possible realizations of the set $\Xc$.
Using (\ref{eq:mse})--(\ref{eq:B}) and the above definitions, 
in Appendix~\ref{app:proof11} we show that, in the case of the LMMSE:
\begin{eqnarray}
{\rm MSE}_{\infty}
&=& \EE_{\lambda_{d,\beta}}\left[\frac{\alpha\beta}{\lambda_{d,\beta}+\alpha\beta}\right]
\label{eq:mseinf_LMMSE3}
\end{eqnarray}
where $\lambda_{d,\beta}$ is a random variable with probability
density function (pdf) $f_{d,\beta}(x)$, distributed as
the asymptotic eigenvalues of $\Tm_d=\beta\Rm_d =\beta\Gm_d\Gm_d^{\dagger}$. 
The subscripts $d$ and $\beta$ of $\lambda$ indicate that the
distribution of the asymptotic eigenvalues of $\Tm_d$ depends on both
the field dimension $d$ and the parameter $\beta$. 

The matrix $\Tm_d$ plays an important role in our analysis; in the following, 
we introduce some of its properties. In the unidimensional case
($d=1$), $\Tm_1$ is a $\left(2M+1\right)\times \left(2M+1\right)$
Hermitian Toeplitz matrix given by 
\[
\Tm_1 =\Tm_1^\dagger = \left( \begin{array}{cccc}
t_0 & t_1 & \cdots & t_{2M} \\
t_{-1} & t_0 & \cdots & t_{2M-1} \\
       &     & \ddots &  \\
t_{-2M}& t_{-2M+1} & \cdots & t_0 
\end{array}\right)
\]
where
$(\Tm_1)_{\ell,\ell'} = t_{\ell-\ell'} =
\frac{1}{r}\sum_{q=1}^{r}\exp(-\jj 2\pi  (\ell-\ell') x_q)$, 
$\ell,\ell'=-M,\ldots,M$. For $d\geq 2$,  $\Tm_d$ can be defined recursively as
a $\left(2M+1\right)^d \times \left(2M+1\right)^d$ Hermitian Block Toeplitz matrix
with non Hermitian Toeplitz blocks:
\begin{eqnarray*}
& &\Tm_d=\left( \begin{array}{cccc}
\Bm_0 & \Bm_1 & \cdots & \Bm_{2M}\\
\Bm_{-1}& \Bm_0&\cdots& \Bm_{2M-1}\\
\vdots & \vdots &  & \vdots\\
\Bm_{-2M}& \Bm_{-2M+1}& \cdots & \Bm_0
\end{array}\right)
\end{eqnarray*}
where
\begin{eqnarray}
(\Tm_d)_{\nu(\ellv),\nu(\ellv')}= \frac{1}{r}\sum_{q=1}^{r} \ee^{-\jj 2 \pi \left(\ellv-\ellv' \right) \xv_q}
\label{t_d}
\end{eqnarray}
and $\ellv, \ellv' \in [-M,\ldots,M]^d$.
That is, the matrix $\Tm_d$ is composed of  $\left(2M+1\right)^2$ blocks $\Bm_i$ of size 
$\left(2M+1\right)^{d-1}\times \left(2M+1\right)^{d-1}$, each 
including $(2M+1)^2$ blocks of size $\left(2M+1\right)^{d-2}\times \left(2M+1\right)^{d-2}$,
and so on. 
The smallest blocks have size $\left(2M+1\right)\times \left(2M+1\right)$; they
have the same structure as matrix $\Tm_1$ in the unidimensional case, however 
only those on the main diagonal have a Hermitian structure. 
Proof of this is given in \cite{Strohmer} for $d=2$; 
the extension to the  $d$-dimensional case is straightforward.


\section{Estimation Error Calculation Method}

The analysis detailed in the next sections consists of the following main steps.
\begin{itemize}
\item[\em{(i)}] As a practical case, we consider the asymptotic expression of the LMMSE in
(\ref{eq:mseinf_LMMSE3}) and notice that 
an analytical evaluation of the asymptotic LMMSE could be 
obtained by exploiting the closed-form expression of the eigenvalue 
distribution, $f_{d,\beta}(x)$, of the reconstruction matrix.
However, such expression is still unknown. Hence, as a first step we derive a closed form expression of the 
moments of $\lambda_{d,\beta}$, for any $d$ and $\beta$, and provide an algorithm
to compute them.
\item[\em{(ii)}] We show that the value of the moments of the eigenvalue distribution
decreases as the field dimension $d$ increases.
\item[\em{(iii)}] We prove that, as $d \rightarrow \infty$, the expression of the
eigenvalue distribution tends to the Mar\v{c}enko-Pastur distribution.
\item[\em{(iv)}] By using the Mar\v{c}enko-Pastur distribution, we are able to 
obtain a tight approximation 
for the LMMSE of the reconstructed field,  which holds for any finite value of $d$.   
\end{itemize}


\section{Closed Form Expression of the moments of the asymptotic eigenvalue pdf}
\label{sec:moments}

Ideally, we would like to obtain the analytical  
expression of the distribution $f_{d,\beta}(x)$ of the asymptotic eigenvalue of $\Tm_d$,
for a given $\beta$. Unfortunately, such a calculation
seems to be prohibitive and is still an open problem. 
Therefore, as a first step, we compute the closed form expression of the moments 
$\EE[\lambda_{d,\beta}^p]$ of $\lambda_{d,\beta}$, for any positive integer $p$.

In the limit for $M$ and $r$ growing to infinity with constant $\beta$,
the expression of $\EE[\lambda_{d,\beta}^p]$ can be easily obtained from 
the powers of $\Tm_d$ as in \cite{Billingsley,Tulino},
\begin{equation}
\EE[\lambda_{d,\beta}^p]
 = \limbeta \frac{1}{(2M+1)^d} \trace \EE_{\Xc}\left[\Tm_d^p\right]
\label{eq:lambda_p}
\end{equation}

In Section~\ref{sec:partitions}, we show that $\EE[\lambda_{d,\beta}^p]$ is a polynomial in $\beta$, of
degree $p-1$ (see (\ref{eq:lambda_p5})); the remaining subsections  
describe how to compute this polynomial.

\subsection{Partitions}
\label{sec:partitions}
Using~(\ref{t_d}), the term
$\trace\EE_{\Xc}[\Tm_d^p]$ in~(\ref{eq:lambda_p}) can be written as
\begin{eqnarray}
&&\hspace{-6ex}\trace\EE_{\Xc}\left[\Tm_d^p\right]
= \EE_{\Xc}\left[\sum_{\ellv_1}(\Tm_d^p)_{\nu(\ellv_1),\nu(\ellv_1)}\right] \non
&=&  \EE_{\Xc}\left[\sum_{\Lm \in
    \Lc_d}(\Tm_d)_{\nu(\ellv_1),\nu(\ellv_2)}\cdots(\Tm_d)_{\nu(\ellv_p),\nu(\ellv_1)}\right]\non 
&=& \frac{1}{r^p} \sum_{\qv \in \Qc}~\sum_{\Lm \in
  \Lc_d}~\EE_{\Xc}\left[\displaystyle\ee^{-\jj 2\pi  \sum_{i=1}^p  
\xv_{q_i}^{\rm T}(\ellv_i-\ellv_{[i+1]})}\right]
\label{eq:lambda_p2}
\end{eqnarray} 
where $\Qc = \{\qv~|~\qv=[q_1, \ldots, q_p]\}$, $q_i=1,\ldots,r$
$\Lc_d = \{\Lm \,|\, \Lm=[\ellv_1, \ldots, \ellv_p]\}$, 
$\ellv_i=\left[\ell_{i_1},\cdots, \ell_{i_d}\right]^{\rm
  T}$, $\ell_{i_m}=-M,\ldots M$
and  
\[ [i+1] = \left\{\begin{array}{ll} i+1 & 1\le i < p \\ 1  & i=p \end{array}\right. \]
In (\ref{eq:lambda_p2}), the average is performed over the random set
of positions $\Xc=\{\xv_1, \ldots, \xv_r\}$, with independent and
uniformly distributed elements. To obtain a closed-form expression of
the distribution moments, we rewrite (\ref{eq:lambda_p2}) by using set
partitioning.

Let $\Pc=\{1,\ldots,p\}$ be the set of integers from 1 to $p$. We
observe that any given vector $\qv \in \Qc$ partitions the set $\Pc$
into $1\le k(\qv)\le p$ disjoint non-empty subsets $\Pc_1(\qv),\ldots,
\Pc_{k}(\qv)$, where $\Pc_j$, $j=1,\ldots,k(\qv)$, is the set of
indices of the entries of $\qv$ taking the same value $\gamma_j$. That
is, 
\begin{equation}
\Pc_j(\qv)= \left\{ i\in \Pc~|~q_i=\gamma_j \right\}
\label{eq:Pj1}
\end{equation}
and $k(\qv)$ is the number of distinct values $\gamma_j$ taken by the
entries of vector $\qv$. Subsets $\Pc_j$ have the following properties
\[ \union_{j=1}^{k(\qv)} \Pc_j(\qv) = \Pc,\,\,\,\,\,\,\,\,\Pc_j(\qv) \cap
\Pc_{j'}(\qv) = \emptyset\,\]
for, $j\neq j'$. Also, we point out that,  since $r$ is the number of
values that the entries $q_i$ can take, there exist $r!/(r-k(\qv))!$ 
vectors $\qv\in \Qc$ generating a given partition of $\Pc$ made of
$k(\qv)$ subsets. In order to clarify the above concepts, we provide
an example below. 

\example{1}{
Let $p=6$, then $\Pc=\{1,2,3,4,5,6\}$. Also, let
$\qv=[4,9,5,5,4,3]$. Since the distinct values in $\qv$ are
$\gamma_j=3,4,5,9$, we have $k(\qv) = 4$. It follows that $\Pc$ is
partitioned into the following subsets: 
\[
\begin{array}{ll} 
\Pc_1(\qv)= \{1,5\} & (q_1=q_5=4)\\
\Pc_2(\qv)= \{2\}   & (q_2=9)\\
\Pc_3(\qv)= \{3,4\} & (q_3=q_4=5)\\ 
\Pc_4(\qv)= \{6\}   & (q_6=3)
\end{array}
\]
Hence, the partition of $\Pc$ induced by $\qv$ is $\{\{1,5\},\{2\},\{3,4\},\{6\}\}$.}

\begin{figure}[ht]
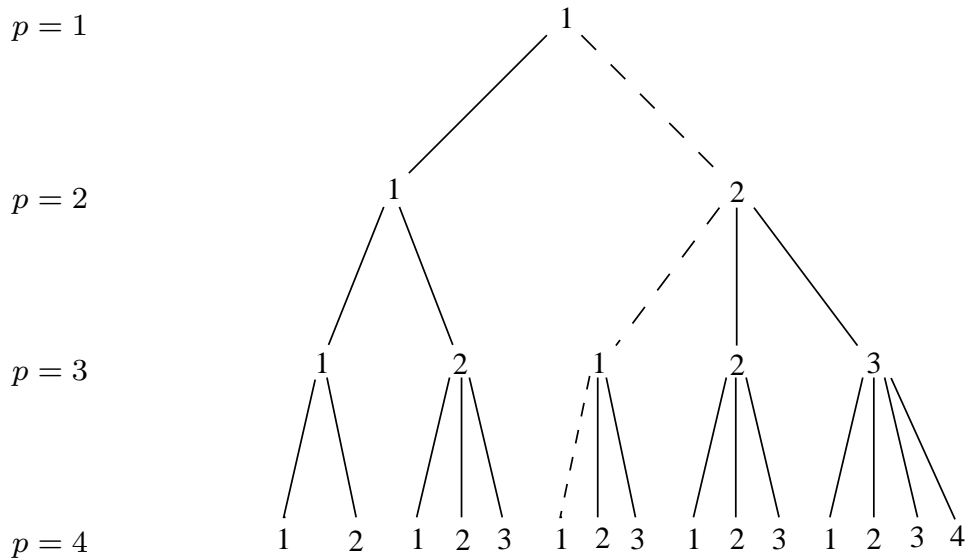

\centerline{\resizebox{0.85\columnwidth}{!}{
\input tree4.pstex_t}}
\caption{Partitions tree of depth
  $p=4$. The path $\omegav=[1,2,1,1]$ employed in Example 2 is
  highlighted by using dashed lines}
\label{fig:tree4}
\end{figure}

Next, we introduce an effective method to represent partitions of a
set $\Pc$, by  building a tree of depth $p$, as in
Figure~\ref{fig:tree4}. Such a representation will allow to simplify
the notation in the following analysis. To build the tree of depth
$p$, we proceed as follows. Each node of the tree is assigned with a
label from the set $\Pc=\{1,\ldots,p\}$, starting from the root which
is labeled by 1. Each node  generates $m+1$ leaves, labeled in
increasing order from 1 to $m+1$, where $m$ is the largest label on
the path from the root to such node. Note that, at level $p$, any
value in $\{1,\ldots,p\}$ is used to label the leaves at least once. 

Then, given a tree of depth $p$, we define  $\omegav= [\omega_1,
  \ldots, \omega_p]$ as a path of length $p$ from the tree's root to a
leaf. We observe that a  vector $\qv$ can be represented as a path
$\omegav$ in the tree of depth $p$. This is done by assigning a label
($j=1,\ldots,p$)  in increasing order to every distinct value of
$\qv$; the vector collecting the labels is the path $\omegav$
corresponding to the given $\qv$. We have that the path $\omegav$,
corresponding to a given $\qv$, defines in the tree of depth $p$ the
same partition  of the set $\Pc$ as the one induced by $\qv$.
Indeed, given a partition of $\Pc$, the subset $\Pc_j$ defined
in~(\ref{eq:Pj1}) can be rewritten as 
\begin{equation}
\Pc_j(\omegav) = \{i\in \Pc\,\mid\,\omega_i = j\},
\label{eq:Pj2}
\end{equation}
i.e., as the set of integers corresponding to the depths of the $j$-th
label in the path. 
As a last remark, consider the number of distinct values in $\qv$
(i.e., the number of distinct labels in $\omegav$) to be equal to
$k(\qv)$, and recall that the number of all possible values taken by
the $p$ elements of $\qv$ is equal to $r$. It follows that
$r!/(r-k(\qv))!$ different $\qv$'s yield the same vector $\omegav$. 
This is in agreement with the fact that, given $\Pc$, there are
$r!/(r-k(\qv))!$ different $\qv$'s generating the same partition
consisting of $k(\qv)$ subsets. Again, for the sake of clarity, we give
an example. 

\example{2}{Let us consider $p=4$ and $\qv=[4, 9, 4, 4]$. 
The vector $\qv$ can be represented in the tree of depth 4 as the path
$\omegav= [1,2,1,1]$ (i.e., the path highlighted with dashed lines in
Figure~\ref{fig:tree4}). In $\qv$ there are two distinct values
(namely, 4 and 9), or, equivalently, in the path $\omegav$ there are
two labels (namely, 1 and 2); then $k(\qv)=k(\omegav)=2$. Label 1
appears in $\omegav$ at depths 1,3, and 4 ($\Pc_1=\{1,3,4\}$), while
label 2 appears at depth 2 ($\Pc_2=\{2\}$). The partition of
$\Pc=\{1,2,3,4\}$ induced by $\qv$ or, equivalently, by $\omegav$ is
therefore: $\{ \{1,3,4\}, \{2\}\}$.} 

From the discussion above, it should be clear that considering a
partition of $\Pc$ is equivalent to considering a path $\omegav$ in a
tree of depth $p$. Hence, in the following analysis, we will  refer to
a partition through its corresponding path $\omegav$. 

We now exploit set partitioning to rewrite (\ref{eq:lambda_p2}). Since
the random vectors $\xv_q$ are independent, given $\qv$, the average
operator in (\ref{eq:lambda_p2}) factorizes into $k(\qv)$ terms, i.e.,
\begin{eqnarray}
&&\EE_{\Xc}\left[\ee^{- \jj 2\pi \sum_{i=1}^p \xv_{q_i}^{\rm
        T}(\ellv_i-\ellv_{[i+1]})}\right] \non
&&\qquad = \prod_{j=1}^{k(\qv)} \EE_{\xv_{\gamma_j}}\left[\ee^{-\jj 2\pi  \xv_{\gamma_j}^{\rm T}
\sum_{i \in \Pc_j(\qv)}\ellv_i-\ellv_{[i+1]}}\right]
\label{eq:factorize1}
\end{eqnarray}
Each term depends on a single random vector
$\xv_{\gamma_j}$. Moreover, since the entries of $\xv_{\gamma_j}$ are
independent random variables uniformly distributed in $[0,1)$, we have: 
\begin{eqnarray}
&&\EE_{\xv_{\gamma_j}}\left[ \ee^{-\jj 2\pi   \xv_{\gamma_j}^{\rm T} 
\sum_{i \in \Pc_j(\qv)}\ellv_i-\ellv_{[i+1]}}\right] \non
&& \quad =\prod_{m=1}^d \EE_{x_{\gamma_j,m}}\left[ \ee^{-\jj 2\pi  x_{\gamma_j,m} c_{jm}} \right]=\prod_{m=1}^d \delta( c_{jm})
\label{eq:factorize2}
\end{eqnarray}
where $x_{\gamma_j,m}$ and $\ell_{i,m}$ are the $m$-th entries of
$\xv_{\gamma_j}$ and $\ellv_i$, respectively, where the function
$\delta(\cdot)$ is the Kronecker's delta, and where
$c_{jm} = \sum_{i \in \Pc_j(\qv)}\ell_{i,m}-\ell_{[i+1],m}$.
By substituting~(\ref{eq:factorize1}) and~(\ref{eq:factorize2}) in
~(\ref{eq:lambda_p2}) and by expanding the summation $\sum_{\Lm \in
  \Lc_d}$, we obtain
\begin{eqnarray}
\trace\EE_{\Xc}[\Tm_d^p]
&=& \frac{1}{r^p}~\sum_{\qv \in \Qc}~\sum_{\ellv_1 \in \Lc_1} \cdots
  \sum_{\ellv_d \in
    \Lc_1}\prod_{j=1}^{k(\qv)}~\prod_{m=1}^{d}~\delta(c_{jm}) \non 
&=& \frac{1}{r^p}~\sum_{\qv \in
    \Qc}~\prod_{m=1}^{d}~\left[~\sum_{\ellv_m \in
      \Lc_1}\prod_{j=1}^{k(\qv)}\delta(c_{jm}) \right] \non 
&=& \frac{1}{r^p}~\sum_{\qv \in \Qc}~\left[~\sum_{\ellv \in \Lc_1}\prod_{j=1}^{k(\qv)}\delta(c_j) \right]^d
\label{eq:lambda_p4}
\end{eqnarray}
where $c_j = \sum_{i \in \Pc_j(\qv)}\ell_i-\ell_{[i+1]}$.
For any given $\qv \in \Qc$, the expression
\begin{eqnarray}
\zeta_M(\qv)=\sum_{\ellv \in \Lc_1}~\prod_{j=1}^{k(\qv)} ~\delta(c_j)
\label{eq:zeta}
\end{eqnarray}
is a polynomial in $M$, since it represents the number of points with
integer coordinates contained in the hypercube $[-M,\ldots,M]^p$ and
satisfying the $k(\qv)$ constraints: 
\begin{equation}
c_j = \sum_{i \in \Pc_j(\qv)}\ell_i-\ell_{[i+1]} = 0
\label{eq:constraints}
\end{equation}
for $j=1,\ldots,k(\qv)$. In Appendix~\ref{app:constraints}, we show  that
one of these constraints is always redundant and that the number of
linearly independent constraints is exactly equal to $k(\qv)-1$. As a
consequence, the polynomial $\zeta_M(\qv)$ has degree $p-k(\qv)+1$,
and, for large values of $M$, we have $\zeta_M(\qv) = O(
(2M+1)^{p-k(\qv)+1} )$.

Now, using (\ref{eq:lambda_p4}) and~(\ref{eq:zeta}), the limit
in~(\ref{eq:lambda_p}) is given by 
\begin{equation}
\EE[\lambda_{d,\beta}^p] = \limbeta~\sum_{\qv \in \Qc}\frac{\zeta_M(\qv)^d}{r^p (2M+1)^d}
\label{eq:moment}
\end{equation}
Equation~(\ref{eq:moment}) can be further simplified by considering
that there exist $r!/(r-k(\qv))!$ vectors $\qv\in \Qc$ generating a
given partition of $\Pc$ made of $k(\qv)$ subsets, or, equivalently, a
path $\omegav$ of length $p$ with $k(\qv)$ distinct labels.

Let $\Omega_p$ be the set of vectors $\omegav$, each corresponding to
a distinct partition of $\Pc$. Also, let us write $k(\qv)$ and
$\zeta_M(\qv)$ as functions of the partition induced by $\qv$, i.e.,
as functions of $\omegav$. Then, we obtain:
\begin{eqnarray}
&&\hspace{-6ex} \EE[\lambda_{d,\beta}^p]
 = \limbeta \sum_{\omegav \in \Omega_p}  \sum_{\qv \Rightarrow
  \omegav} \frac{\zeta_M(\omegav)^d }{r^p (2M+1)^d} \non 
&\overset{(a)}{=}& \limbeta \sum_{\omegav \in \Omega_p} \frac{
  \zeta_M\left(\omegav\right)^d r!}{(r-k(\omegav))!\,r^p\,(2M+1)^d}
\label{eq:lambda_p3a}
\end{eqnarray}
where 
\begin{itemize}
\item the notation $\sum_{\qv \Rightarrow \omegav}$ represents
the sum over all vectors $\qv$ generating a given path $\omegav$,
\item the equality $(a)$ holds because the number of vectors $\qv$ generating a
given $\omegav$ is $r!/(r-k(\omegav))!$.
\end{itemize}
Note that, for large $r$, $r!/(r-k(\omegav))! = r^{k(\omegav)} +O(r^{k(\omegav)-1})$. 
Also, since $\zeta_M\left(\omegav\right)$ is a polynomial
in $M$ of degree $p-k(\omegav)+1$, for large values of $M$ we have:
$\zeta_M\left(\omegav\right) =  v(\omegav) (2M+1)^{p-k(\omegav)+1}+ O((2M+1)^{p-k(\omegav)})$,
where $v(\omegav)$ is the coefficient of degree $p-k(\omegav)+1$ of $\zeta_M(\omegav)$.
Therefore, taking the limit, we obtain:
\begin{eqnarray}
 \EE[\lambda_{d,\beta}^p] 
= \sum_{\omegav \in \Omega_{p}} v(\omegav)^d
   \beta^{p-k(\omegav)} 
= \sum_{k=1}^p \beta^{p-k} \sum_{\omegav \in \Omega_{p,k}} v(\omegav)^d
\label{eq:lambda_p5}
\end{eqnarray}
where $\Omega_{p,k}\subseteq \Omega_p$ is the subset of $\Omega_p$ 
containing paths with $k(\omegav)$ distinct labels,
and%
\begin{equation}
v(\omegav) = \lim_{M \rightarrow +\infty} \frac{\zeta_M(\omegav)}{(2M+1)^{p-k(\omegav)+1}}
\label{eq:v}
\end{equation}
Note  that the coefficient $v(\omegav)$ represents
the volume of the {\em convex polytope} described by the constraints in 
(\ref{eq:constraints}),
when the variables $\ell_i$ are considered as real and limited to
a $p$-dimensional hypercube of volume $1$. 
As a consequence, we have:  $0 \le v(\omegav) \le 1$.

Equation (\ref{eq:lambda_p5}) provides a closed-form expression of the moment $\EE[\lambda_{d,\beta}^p]$, as 
a polynomial in $\beta$ of degree $p-1$. Again, for the sake of clarity, we give an example below. 

\example{3}{Let $p=6$ and $\qv=[4,9,5,5,4,3]$. We have $\omegav=[1,2,3,3,1,4]$, and
the partition of $\Pc$ is $\{ \{1,5\}, \{2\}, \{3,4\}, \{6\}\}$.
Then, the set of $k(\omegav)=4$ constraints (as in (\ref{eq:constraints})) are given by:
\begin{eqnarray*}
\ell_1 + \ell_5 = \ell_2 + \ell_6, && \ell_2 = \ell_3, \\
\ell_3 + \ell_4 = \ell_4+ \ell_5, && \ell_6 = \ell_1 
\end{eqnarray*}
The last equation is redundant since can be obtained from the first three constraints.
Simplifying, we obtain $\ell_1=\ell_6$, and $\ell_2=\ell_3=\ell_5$.
Since each variable $\ell_i$ ranges from $-M$ to $M$, the number of integer solutions
satisfying the constraints is exactly $\zeta_{M}(\omegav)=(2M+1)^3$, and then 
$v(\omegav)=\lim_{M \rightarrow +\infty} \frac{\zeta_M(\omegav)}{(2M+1)^{p-k(\omegav)+1}}=1$.} 

Next, in order to compute $\EE[\lambda_{d,\beta}^p]$, we need:
\begin{itemize}
\item to enumerate the partitions, i.e., the vectors $\omegav \in
  \Omega_{p,k}$, for each $k=1,\ldots,p$  (see
  Section~{\ref{sec:partition_enumeration}}); 
\item to compute the coefficients $v(\omegav)$, for any $\omegav \in
  \Omega_{p,k}$ and $k=1,\ldots,p$ (see
  Section~{\ref{sec:computation_vtau}}).
\end{itemize}

\subsection{Partitions enumeration}
\label{sec:partition_enumeration}
We notice that $\Omega_p$ represents the set of partitions of a
$p$-element set, thus it has cardinality $|\Omega_p| = B(p)$, where
$B(p)$ is the $p$-th {\em Bell number} or {\em exponential
  number}~\cite{bellnumbers}. Furthermore,  the subset $\Omega_{p,k}
\subseteq \Omega_p$ has cardinality $S(p,k)$, which is a  
{\em Stirling number of the second kind}~\cite{stirling2numbers} given
by: 
\[ S(p,k) = \frac{1}{k!}\sum_{i=0}^k (-1)^i {k\choose i}(k-i)^p \]
with $B(p) = \sum_{k=1}^p  S(p,k)$.

\subsection{Computation of the coefficients $v(\omegav)$}
\label{sec:computation_vtau}
The last step required for the computation of
$\EE[\lambda_{d,\beta}^p]$ is the evaluation of the coefficients
$v(\omegav)$, for every $\omegav \in \Omega_p$.
We have the following Lemma:
\begin{lemma}
\label{th:volume}
For any $\omegav \in \Omega_p$ (or, equivalently, any partition of $\Pc$) 
and any arbitrary integer $n$, with $n=1,\ldots,k(\omegav)$, the
coefficient $v(\omegav)$ in~(\ref{eq:v}) is given by:
\begin{equation}
v(\omegav) = \int_{\RR^{k(\omegav)-1}} \prod_{i=1}^p \sinc \left(y_{\omega_i}-y_{\omega_{[i+1]}}\right)\mid_{y_n=0}\dd \yv_n
\label{eq:volume}
\end{equation}
where 
\begin{itemize}
\item $\yv_n=[y_1,\ldots,y_{n-1},y_{n+1},\ldots,y_{k(\omegav)}]^{\rm T}$,
\item $\sinc(y)= \displaystyle\frac{\sin(\pi y)}{\pi y}$.
\end{itemize}
\end{lemma}
\begin{IEEEproof}
The proof can be found in Appendix~\ref{eq:proof_lemma}.
\end{IEEEproof}

\subsection{A practical method for the moments computation}

Equation (\ref{eq:volume}) in Section~\ref{sec:computation_vtau} shows
that the computation of $\EE[\lambda_{d,\beta}^p]$ requires the
evaluation of $B(p)$ integrals. However, $B(p)$ is very large even for
small $p$, e.g., $B(10)=115975$ and $B(20)\approx 5\cdot 10^{13}$.

The computational complexity can be reduced by recursively applying the 
simplification rules defined in the following Lemma:
\begin{lemma}
\label{th:reduction_rules}
Let 
\begin{itemize}
\item $\omegav=[\omega_1, \ldots, \omega_p]$ be the path in a tree of depth $p$, 
corresponding to the partition of $\Pc$ into $k$ subsets;
\item $\Pc_1, \ldots, \Pc_k$ be the subsets of $\Pc=\{1,\ldots,p\}$ defined as in~(\ref{eq:Pj2});
\item $i \in \Pc_j$:
\item $\omegav'$ be the path obtained from $\omegav$ by removing $\omega_i$.
\end{itemize}
We have the following rules: 
\begin{enumerate}
\item if $\Pc_j$ has cardinality 1 (i.e., $\Pc_j$ is a {\em singleton}) or
\item if $\Pc_j$ contains adjacencies (in the circular sense), i.e., both $i$ and $[i+1] \in \Pc_j$,
\end{enumerate}
then $v(\omegav) = v(\omegav')$
\end{lemma}
\begin{IEEEproof}
The proof is a direct consequence of Lemma~\ref{th:volume} and can be found in Appendix~\ref{app:reduction_rules}.
\end{IEEEproof}
Table~\ref{tab:reduction examples} shows two examples of how the rules described in Lemma~\ref{th:reduction_rules}
can be applied.
Example 1 in the Table assumes $p=6$ and  $\omegav=[1,2,3,2,2,1]$. At step 1, we note that
the third element  ($i=3$) of $\omegav$ is a singleton, then, by applying rule 1, we can 
remove it from the path.
At step 2, we find that in $\omegav$ there are some adjacencies, hence we apply twice rule 2 (steps 2 and 3).
At step 4, the second element of $\omegav$ is a singleton, and we remove it by applying rule 1.
Eventually, at step 7, the path $\omegav$ is empty (i.e., has size $p=0$) and, thus, the corresponding 
coefficient is $v(\omegav)=1$ $(\EE[\lambda_{d,\beta}^0]=\EE[1]=1)$.
 
Example 2 in the Table assumes $p=6$ and $\omegav=[1,2,3,1,2,1]$. After removing a singleton
(step 1) and an adjacency (step 2), the remaining path cannot be further reduced. Then, to compute the coefficient $v(\omegav)$,
we need to apply directly Lemma~\ref{th:volume} on the path $\omegav=[1,2,1,2]$.
We obtain: 
\begin{eqnarray}
v(\omegav) &=& \int_{-\infty}^{+\infty}  {\rm sinc}(y_1-y_2)\,{\rm
  sinc}(y_2-y_1) \non
&& \qquad \cdot {\rm sinc}(y_1-y_2)\,{\rm sinc}(y_2-y_1)\mid_{y_2=0} \dd y_1\non 
&=&\int_{-\infty}^{+\infty} {\rm sinc}(y_1)^4 \dd y_1 = \frac{2}{3} \nonumber
\end{eqnarray}

\begin{table}
\caption{Example of complexity reduction using the rules described in Lemma~\ref{th:reduction_rules} }
\begin{center}
\begin{tabular}{||c||c|c|c||c|c|c||} \hline
 &\multicolumn{3}{c||}{Example 1} &\multicolumn{3}{c||}{Example 2}\\ \hline 
Step & $\omegav$ & Rule & $i$ & $\omegav$ & Rule & $i$\\ \hline
1 & [1,2,3,2,2,1] & 1 & 3 & [1,2,3,1,2,1] & 1 & 3\\ \hline
2 &[1,2,2,2,1] & 2 & 2 & [1,2,1,2,1] & 2 & 5\\ \hline
3 &[1,2,2,1] & 2 & 2 & [1,2,1,2] &  & \\ \hline
4 &[1,2,1] & 1 & 2 &  &  &\\ \hline
5 &[1,1] & 2 & 2   &  &  &\\ \hline
6 &[1] & 1 & 1     &  &  &\\ \hline
7 &[] &  &         &  &  &\\ \hline
&\multicolumn{3}{c||}{$v(\omegav) = 1$} &\multicolumn{3}{c||}{$v(\omegav) = 2/3$}\\ \hline 
\end{tabular}
\end{center}
\label{tab:reduction examples}
\end{table}

In the following example, Lemmas~\ref{th:volume} and~\ref{th:reduction_rules} are 
exploited to explicitly compute $\EE[\lambda_{d,\beta}^4]$.

\example{4}{Let us consider $p=4$.
The total number of partitions of $\Pc=\{1,2,3,4\}$ is equal to $B(4)=15$.
Considering the tree of depth $p$, we apply to each path the rules of 
Lemma~\ref{th:reduction_rules}, and we find that 
14 paths (partitions) out of 15 reduce to the
empty path, thus contributing with $v(\omegav)=1$. 
The only path that cannot be further reduced is $\omegav=[1,2,1,2]$. Thus, 
applying Lemma~\ref{th:volume} with $n=2$, we obtain $v(\omegav)=2/3$.
From (\ref{eq:lambda_p5}) and considering all contributions, we obtain:
\[ \EE[\lambda_{d,\beta}^4] = \beta^3 + \left(6+ (2/3)^d\right)\beta^2 + 6\beta + 1.\]
}

\section{Convergence to the Mar\v{c}enko-Pastur distribution}
\label{sec:approx-distribution}

In Section~\ref{sec:moments}, we have shown that the moments of the
asymptotic eigenvalues of $\Tm_d$ are polynomials in $\beta$, given
by~(\ref{eq:lambda_p5}). In particular, the $p$-th moment
$\EE[\lambda^p_{d,\beta}]$ has degree $p-1$ and is given by the sum of
$B(p)$ positive contributions of the form $v(\omegav)^d
\beta^{p-k(\omegav)}$. Since $0<v(\omegav)\le 1$ and $\beta>0$, for
any $d$, the following inequality holds: 
\[  \EE[\lambda^p_{d+1,\beta}] \le  \EE[\lambda^p_{d,\beta}] \]
i.e., for any given $p$ and $\beta$, the moments of the asymptotic
eigenvalues decrease as the field dimension increases. The series
$\EE[\lambda^p_{d,\beta}]$, as a function of $d$, is positive and
monotonically decreasing, thus it converges to:
\begin{equation}
   \EE[\lambda^p_{\infty,\beta}] = \lim_{d \rightarrow +\infty } \EE[\lambda^p_{d,\beta}]
\end{equation}

\begin{lemma}
\label{th:mp}
The moments $\EE[\lambda^p_{\infty,\beta}]$ are the Narayana
polynomials, given by 
\begin{equation}
\EE[\lambda^p_{\infty,\beta}] = \sum_{k=1}^p T(p,k) \beta^{p-k}
\end{equation}
where $T(p,k)= \frac{1}{k}\binom{p-1}{k-1}\binom{p}{k-1}$ are the {\em
  Narayana numbers}~\cite{EncIntSeq,Dumitriu}. Moreover, the random
variable $\lambda^p_{\infty,\beta}$ follows the Mar\v{c}enko-Pastur
distribution~\cite{MarcenkoPastur} with pdf (see Figure
\ref{fig:mar-pas}): 
\begin{equation}
 f_{\infty,\beta}(x) = \frac{\sqrt{(c_1-x)(x-c_2)}}{2\pi x \beta}
\end{equation}
where $c_1,c_2 = (1 \pm \sqrt{\beta})^2$, $0 < \beta \le 1$, $c_2 \le x \le c_1$.
\end{lemma}
\begin{IEEEproof}
The proof is given in Appendix~\ref{app:mp}.
\end{IEEEproof}

\insertfig{0.85}{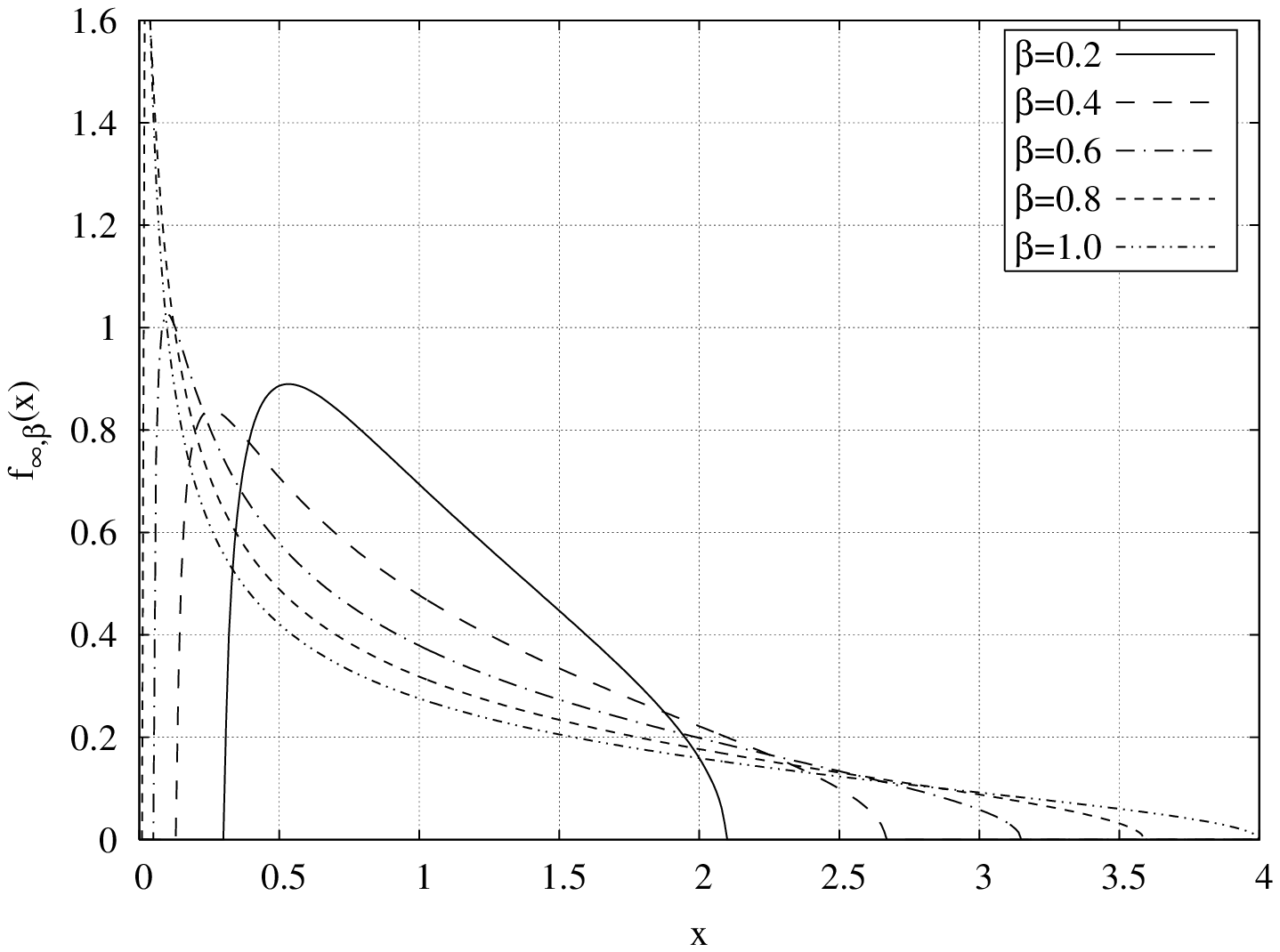}{Mar\v{c}enko-Pastur
  distribution}{fig:mar-pas}

In the following, we apply our findings to the study of the LMMSE of a
reconstructed multidimensional field; in particular, we exploit the
Mar\v{c}enko-Pastur distribution to compute the expectation
in~(\ref{eq:mseinf_LMMSE3}).

\insertfig{0.85}{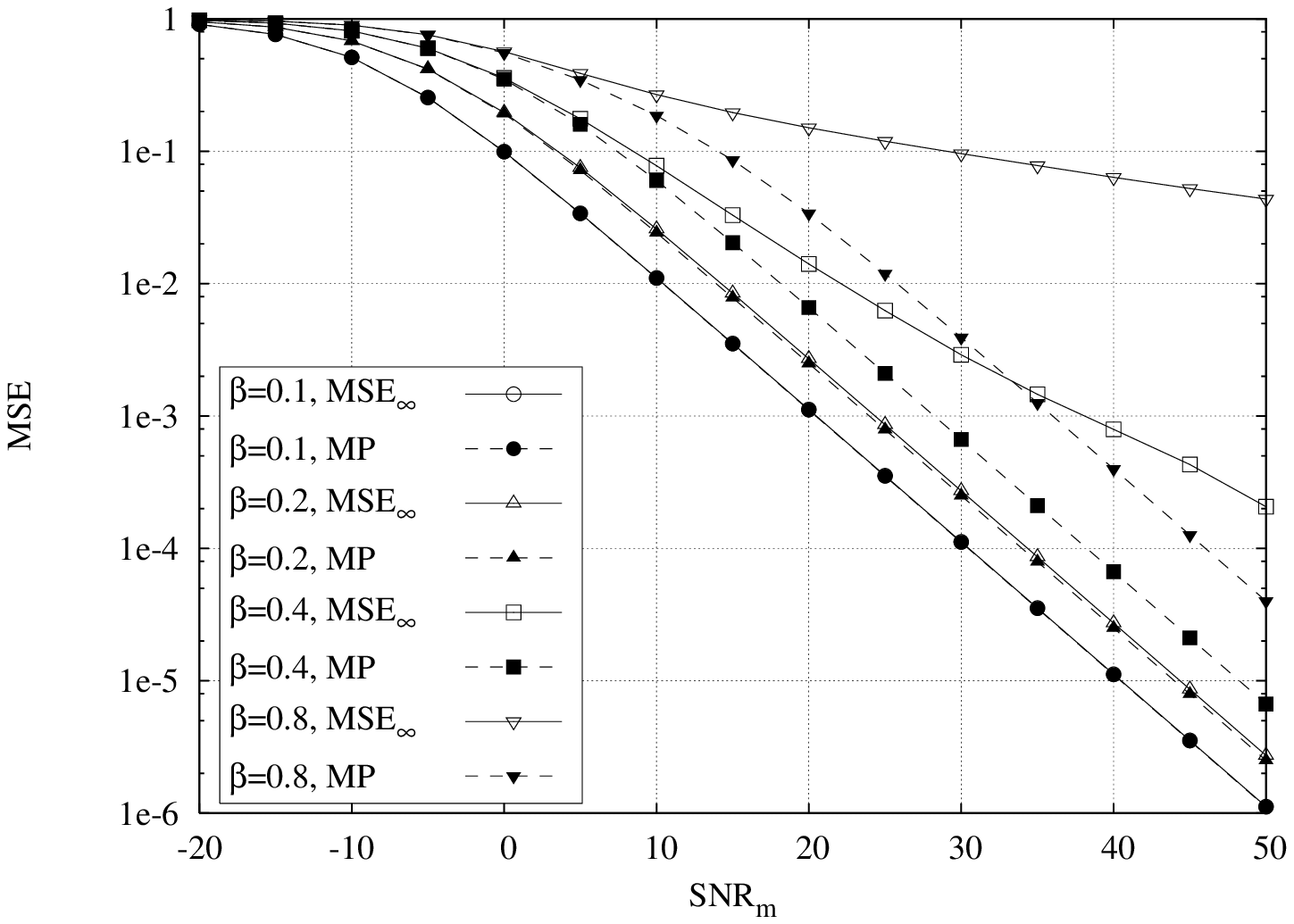}{MSE of the reconstructed field for
  $d=1$ and varying values of $\beta$. Comparison between the MSE
  asymptotic value (\ref{eq:mseinf_LMMSE3}) and the fully analytical
  expression derived using the Mar\v{c}enko-Pastur distribution
  (\ref{eq:MSE_MP})}{fig:MSEunifvarBeta}  
\insertfig{0.85}{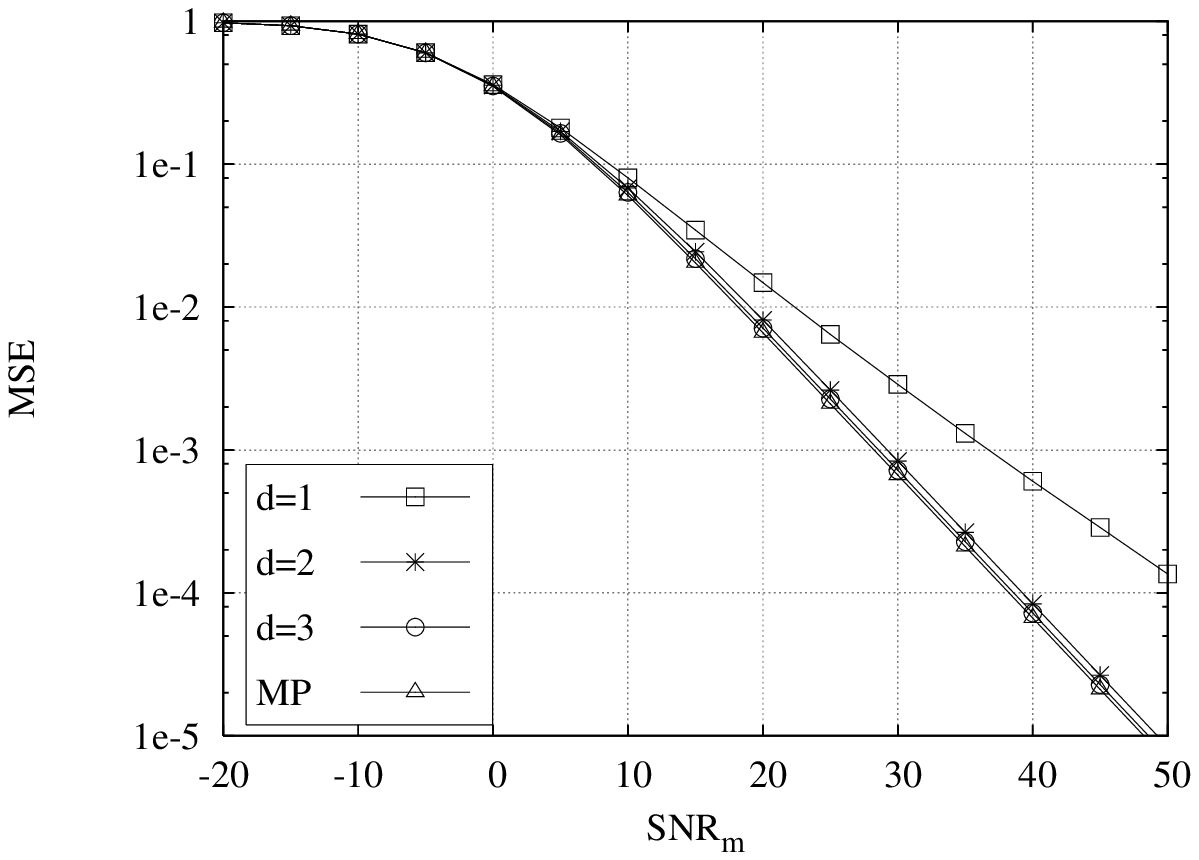}{MSE of the reconstructed field,
  for $\beta=0.4$ and $d=1,2,3$. Comparison between the MSE asymptotic
  value (\ref{eq:mseinf_LMMSE3}) and the fully analytical expression
  derived using the Mar\v{c}enko-Pastur distribution
  (\ref{eq:MSE_MP})}{fig:MSEunifvardBeta04} 
\insertfig{0.85}{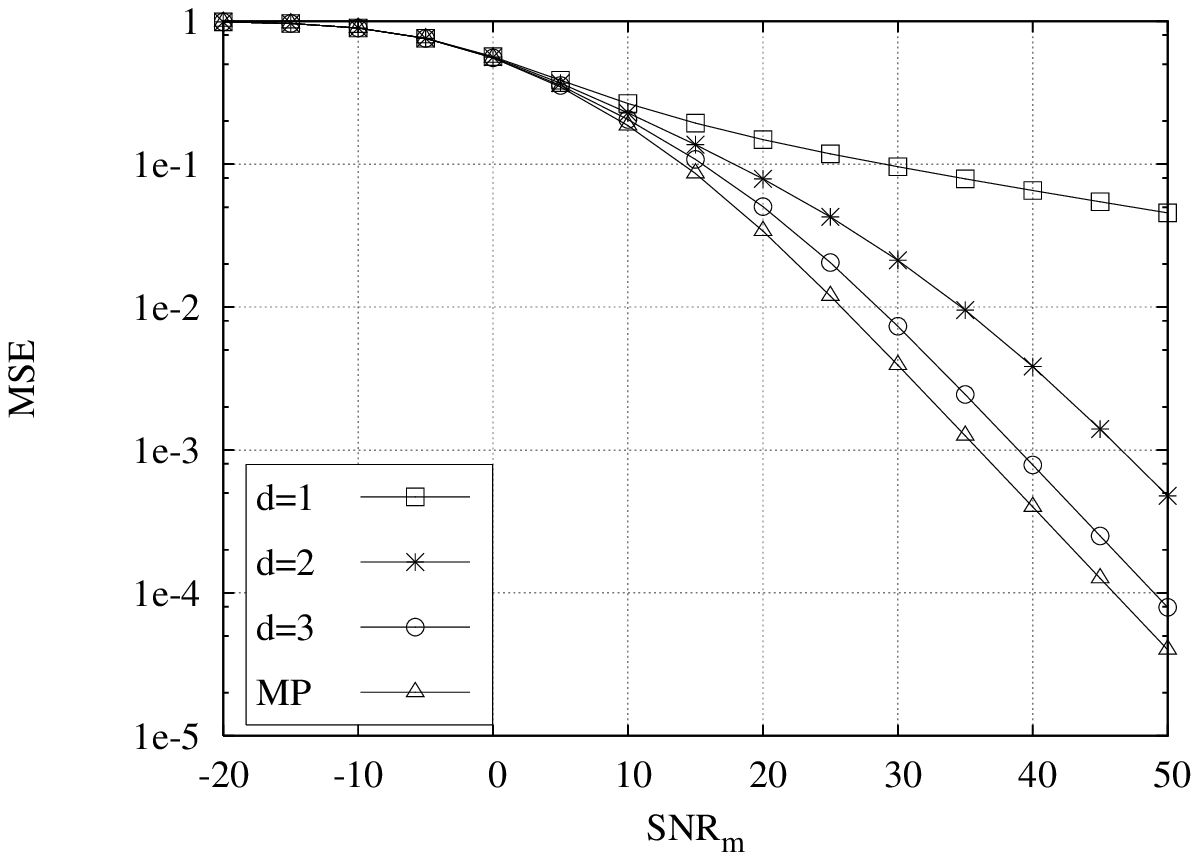}{MSE of the reconstructed field,
  for $\beta=0.8$ and $d=1,2,3$. Comparison between the MSE asymptotic
  value (\ref{eq:mseinf_LMMSE3}) and the fully analytical expression
  derived using the Mar\v{c}enko-Pastur distribution
  (\ref{eq:MSE_MP})}{fig:MSEunifvardBeta08}

\section{Study of the reconstruction quality through the Mar\v{c}enko-Pastur distribution}
\label{sec:exploitation}

Recall that the MSE provided by the LMMSE filter is~\cite{infocom07}:
\begin{equation*}
{\rm MSE}_{\infty}^{\rm LMMSE} =
\EE_{\lambda_{d,\beta}}\left[\frac{\alpha\beta}{\lambda_{d,\beta}+\alpha\beta}\right]
\end{equation*}
where $\lambda_{d,\beta}$ is distributed as the asymptotic eigenvalues
of $\Tm_d$, with pdf $f_{d,\beta}(x)$.

By using the Mar\v{c}enko-Pastur distribution $f_{\infty,\beta}$
instead of $f_{d,\beta}$, we have: 
\begin{eqnarray}
&&\hspace{-6ex}{\rm MSE}_{\infty} = \EE_{\lambda_{\infty,\beta}}\left[\frac{\alpha\beta}{\lambda_{\infty,\beta}+\alpha\beta}\right]
= \int_{c_2}^{c_1} \frac{\alpha\beta f_{\infty,\beta}(x)}{x+\alpha\beta} \dd x \non
&=& \frac{2\beta-\theta+\sqrt{\theta^2-4\beta}}{2\beta} 
\label{eq:MSE_MP}
\end{eqnarray}
where $\theta = 1+\beta(1+\alpha)$.

Equation (\ref{eq:MSE_MP}) provides an approximation to the ${\rm MSE}_{\infty}$, 
which, as 
shown in the following plots, can be exploited to derive the quality of the reconstructed field, given a finite $d$. 

We first consider $d=1$ and compare in Figure \ref{fig:MSEunifvarBeta}
the expression of the ${\rm MSE}_{\infty}$ 
as in (\ref{eq:mseinf_LMMSE3}) (solid lines) with the one obtained by 
using the Mar\v{c}enko-Pastur  distribution  (dashed lines).
The results are presented as functions of the SNR$_m$ and for different values of $\beta$.
We computed (\ref{eq:mseinf_LMMSE3})
by averaging over the eigenvalues of 200 realizations  of the matrix
$\Tm_1$, with $M=150$.
The plot shows that, for small values of $\beta$, 
the Mar\v{c}enko-Pastur distribution (\ref{eq:MSE_MP})
yields an excellent approximation to the ${\rm MSE}_{\infty}$, already for $d=1$.
Instead, for  values of $\beta$ greater than 0.2, the expression
in (\ref{eq:MSE_MP}) fails to provide a valid approximation.

However, it is interesting to notice that, for $d>1$, it is possible to obtain
an accurate approximation of the ${\rm MSE}_{\infty}$ using the Mar\v{c}enko-Pastur distribution, 
even for large values of $\beta$.
This is shown by Figures \ref{fig:MSEunifvardBeta04} and \ref{fig:MSEunifvardBeta08},
which plot the results obtained through (\ref{eq:mseinf_LMMSE3}) and 
(\ref{eq:MSE_MP}) for $\beta$ equal to 0.4 and 0.8,  respectively. 
The results are presented as the SNR$_m$ varies and for different values of the field dimension $d$.

Looking at Figure~\ref{fig:MSEunifvardBeta04}, 
we note that our approximation is tight for $d\ge 2$, while 
Figure~\ref{fig:MSEunifvardBeta08} shows 
that, when $d=3$, we still 
get a fairly good approximation for $\beta$ as large as $0.8$.

\section{Conclusions}
\label{sec:conclusions}

We considered a large-scale wireless sensor network sampling a
multidimensional field, and we investigated the mean square error
(MSE) of the signal reconstructed at the sink node. We noticed that
an analytical study of the quality of the reconstructed field could be
carried out by using the eigenvalue distribution of the matrix
representing the sampling system. Since such a distribution is unknown, 
we first derived a closed-form
expression of the distribution moments. By using this expression, we
were able to show that the eigenvalue distribution of the
reconstruction matrix tends to the Mar\v{c}enko-Pastur distribution as
the field dimension tends to infinity. We applied our results to the
study of the MSE of the reconstructed field, when linear filtering is
used at the sink node. We found that, by using the Mar\v{c}enko-Pastur
distribution instead of the actual eigenvalue distribution, we obtain
a close approximation to the MSE of the reconstructed signal, which
holds for field dimensions $d \geq 2$.

We believe that our work is the basis for an analytical study of
various aspects concerning the reconstruction quality of
multidimensional sensor fields, and, more generally, of irregularly
sampled signals.

\appendices

\section{The constraints}
\label{app:constraints}
Let us consider a vector of integers $\qv$ of size $p$ 
partitioning the set $\Pc=\{1,\ldots,p\}$ in $k$ subsets
$\Pc_j$, $1\le j \le k$, and the set of $k$
constraints~(\ref{eq:constraints}). We first show that one of these
constraints is always redundant. 

\subsection{Redundant constraint}
Choose an integer $j$, $1\le j \le k$. Summing up all constraints
except for the $j$-th, we get:
\[ \sum_{\substack{h=1\\ h\neq j}}^k c_h =\sum_{h=1}^k c_h - c_j 
  = \sum_{i \in \Pc} \ell_i-\ell_{[i+1]} -  c_j = - c_j \]
which gives the $j$-th constraint since $\sum_{i \in \Pc}
\ell_i-\ell_{[i+1]}=0$. Thus, one of the constraints in
(\ref{eq:constraints}) is always redundant. Next, we show that the
remaining $k-1$ constraints are linearly independent. 

\subsection{Linear independence}
The $k$ constraints in (\ref{eq:constraints}) can be arranged in the
form: $\Wm\ellv =  \zerov $ with $\ellv=[\ell_1,\ldots,\ell_p]^{\rm
  T}$ and $\Wm$ being a $k\times p$ matrix defined as
\begin{equation}
\Wm  = \Wm' - \Wm''
\label{eq:A}
\end{equation}
where
\[(\Wm')_{j,i} = \left\{ \begin{array}{ll} +1 & i\in \Pc_j \\ 0 & {\rm otherwise}
\end{array} \right. \]
and $\Wm''$ is obtained from $\Wm'$ by circularly shifting the
rows to the right by one position. Since one of the
constraints~(\ref{eq:constraints}) is redundant, the rank of $\Wm$ is: 
$\rho(\Wm)\le k-1$. Now we prove that the rank of $\Wm$ is equal to $k-1$.

Since the subsets $\Pc_j$ have empty intersection, the rows of $\Wm'$
are linearly independent; hence, $\Wm'$ has rank $k$. Also, $\Wm''$ is
obtained from $\Wm'$ by circularly shifting the rows by one
position to the right, thus $\Wm''$ can be written as $\Wm'' =
\Wm'\Sm$ where $\Sm$ is the $p\times p$ {\em right-shift
  matrix}~\cite{Brookes}, i.e., the entries of the $i$-th row of $\Sm$
are zeros except for an entry equal to  1 at position $[i+1]$. As a
consequence, $\Wm = \Wm' - \Wm'\Sm = \Wm'(\Id_p - \Sm)$ where
the rows of the matrix $\Id_p - \Sm$ are obtained by circularly
shifting the vector $[+1, -1, 0, \ldots, 0]$ and thus 
has rank $\rho(\Id_p - \Sm)=p-1$. Hence, using the properties of the
rank of matrix products reported in~\cite{Brookes}, we have  
\[\rho(\Wm) = \rho(\Wm'(\Id_p - \Sm)) \ge  \rho(\Wm')+\rho(\Id_p -
\Sm) -p = k-1\]
We recall that the system of linear equations $\Wm\zv$
has a finite number of integer solutions
bounded in $[-M,\ldots, M]^d$.
The number of solutions decreases as $\rho(\Wm)$ increases.

\section{Proof of Lemma~\ref{th:volume}}
\label{eq:proof_lemma}

\begin{IEEEproof}
 Using~(\ref{eq:zeta})
and~(\ref{eq:v}), we obtain:
\[ v(\omegav) = \lim_{M \rightarrow +\infty}
\frac{1}{(2M+1)^{p-k(\omegav)+1}} \sum_{\ellv \in
  \Lc_1}~\prod_{j=1}^{k(\omegav)}~\delta( c_j) \] 
We first notice that $\prod_{j=1}^{k(\omegav)}~\delta(c_j) = \delta( \Wm\ellv)$
where the $k(\omegav) \times p$ matrix $\Wm$ is defined in
Appendix~\ref{app:constraints} and $\delta( \Wm\ellv)$ is 
a multidimensional Kronecker delta \cite{MDKronecker}. Since the rank
of $\Wm$ is $\rho(\Wm)=k(\omegav)-1$ (see
Appendix~\ref{app:constraints}), 
then $\delta( \Wm\ellv)$ defines a subspace of $\ZZ^p$ with $p-k(\omegav)+1$ dimensions.
Therefore, considering that $\ellv$ is a vector of integers with
entries ranging in the interval $[-M,\ldots,M]$ 
and taking the limit for $M\rightarrow \infty$, we obtain
$v(\omegav) = \int_{[-1/2,1/2)^p} \delta_d\left( \Wm\zv\right) \dd \zv$
where $\zv\in \RR^p$ and the function $\delta_d$ represents the Dirac
delta. We have that $\delta_d(\Wm\zv)$ can be factorized as
\begin{equation}
\delta_d(\Wm\zv) = \prod_{j=1}^{k(\omegav)}\delta_d(\rv_j\Tran\zv)
\label{eq:delta}
\end{equation}
where $\rv_j\Tran$ is the $j$-th row of $\Wm$. 
As already shown in Appendix~\ref{app:constraints}, one of the
constraints~(\ref{eq:constraints}) is redundant and, hence, one of the
factors in the right hand side of~(\ref{eq:delta}), say the $n$-th,
must not be included in the product. Now, moving  to the Fourier
transform domain, we can write: 
$\delta_d(\rv_j\Tran\zv) = \int_{-\infty}^{+\infty}\exp(\jj 2\pi  y
\rv_j\Tran\zv) \dd y$. Therefore,
\begin{eqnarray*}
v(\omegav)&=& \int_{[-1/2,1/2)^p} \prod_{j=1,j\neq n}^{k(\omegav)}\delta_d(\rv_j\Tran\zv)\dd \zv \non
&=& \int_{[-1/2,1/2)^p} \prod_{j=1,j\neq n}^{k(\omegav)} \int_{-\infty}^{+\infty}\ee^{\jj 2\pi  y_j \rv_j\Tran\zv} \dd y_j \dd \zv \non 
&=& \int_{[-1/2,1/2)^p}  \int_{\RR^{k-1}} \ee^{\jj 2\pi \sum_{j=1,j\neq n}^{k(\omegav)} y_j \rv_j\Tran\zv} \dd \yv_n \dd \zv 
\end{eqnarray*}
where $\yv_n = [y_1, \ldots, y_{n-1}, y_{n+1}, \ldots, y_{k(\omegav)}]^{\rm T}$.
Integrating first with respect to $\zv$, we get
\begin{eqnarray*}
v(\omegav)
&=& \int_{\RR^{k-1}} \int_{[-1/2,1/2)^p} \ee^{\jj 2\pi \sum_{i=1}^p z_i \yv_n^{\rm T} \wv_i} \dd \zv \dd \yv_n \non 
&=& \int_{\RR^{k-1}} \int_{[-1/2,1/2)^p} \prod_{i=1}^p \ee^{\jj 2\pi z_i \yv_n^{\rm T} \wv_i} \dd \zv \dd \yv_n \non 
&=& \int_{\RR^{k-1}} \prod_{i=1}^p \int_{-1/2}^{1/2} \ee^{\jj 2\pi z_i \yv_n^{\rm T} \wv_i} \dd z_i \dd \yv_n \non 
&=& \int_{\RR^{k-1}} \prod_{i=1}^p \ee^{\jj\pi \yv_n^{\rm T} \wv_i}\, \sinc(\yv_n^{\rm T} \wv_i) \dd \yv_n \non
&=& \int_{\RR^{k-1}} \ee^{\jj\pi  \yv_n^{\rm T} \sum_{i=1}^p \wv_i} \prod_{i=1}^p \sinc(\yv_n^{\rm T} \wv_i) \dd \yv_n 
\end{eqnarray*}
where $\wv_i$ is the $i$-th column of $\Wm$, taken after removing its
$n$-th row. By definition, the $j$-th rows of $\Wm'$ and of $\Wm''$
contain both $p-|\Pc_j|$ ``0'' and $|\Pc_j|$ ``+1''. Since $\Wm = \Wm'
- \Wm''$, we  have $\sum_{i=1}^p \wv_i = \zerov$ and $v(\omegav) =
\int_{\RR^{k-1}} \prod_{i=1}^p \sinc(\yv_n^{\rm T} \wv_i) \dd
\yv_n$. Notice that, by definition of $\Wm$ (see~(\ref{eq:A})),
$\yv_n^{\rm T} \wv_i = y_j - y_{j'} \mid_{y_n=0}$ if $i\in \Pc_j$ and
$[i+1] \in \Pc_{j'}$. Moreover, by the definition in (\ref{eq:Pj2}),
we have $y_j = y_{\omega_i}$ when  $i\in \Pc_j$. Thus, 
\[ v(\omegav) = \int_{\RR^{k-1}} \prod_{i=1}^p \sinc(y_{\omega_i}-y_{\omega_{[i+1]}})\mid_{y_n=0} \dd \yv_n \]
\end{IEEEproof}

\section{Proof of Lemma~\ref{th:reduction_rules} (Simplification Rules)}
\label{app:reduction_rules}
Let $\Pc_j$ be a singleton with $\Pc_j={i}$ and $\omega_i=j$. We first notice that,
since $\Pc_j$ is a singleton, $[i-1],[i+1] \notin \Pc_j$.
By applying Lemma~\ref{th:volume} with an arbitrary $n\neq j$, we have
\begin{eqnarray*}
&&\hspace{-6ex} v(\omegav)= \int_{\RR^{k-1}} \prod_{h=1}^p \sinc(y_{\omega_h}-y_{\omega_{[h+1]}})\mid_{y_n=0} \dd \yv_n \non
&=& \int_{\RR^{k-1}} \prod_{\substack{h=1 \\ h\neq i \\ h \neq
    [i-1]}}^p \sinc(y_{\omega_h}-y_{\omega_{[h+1]}}) \sinc(y_{\omega_{[i-1]}}-y_{\omega_{i}})\non
&&\qquad \cdot \sinc(y_{\omega_i}-y_{\omega_{[i+1]}}) \mid_{y_n=0}  \dd \yv_{n}
\end{eqnarray*}
We now integrate with respect to $y_j$, with $j=\omega_i$ and we obtain
\begin{eqnarray*}
&&\hspace{-6ex} v(\omegav) = \int_{\RR^{k-2}} \prod_{\substack{h=1 \\ h\neq i \\ h \neq
    [i-1]}}^p \sinc(y_{\omega_h}-y_{\omega_{[h+1]}})  \non
&& \cdot \int_{\RR}\sinc(y_{\omega_{[i-1]}}-y_j)\sinc(y_j-y_{\omega_{[i+1]}}) \mid_{y_n=0} \dd y_j \dd \yv'_n \non
&=& \int_{\RR^{k-2}} \prod_{\substack{h=1 \\ h\neq i \\ h \neq
      [i-1]}}^p \sinc(y_{\omega_h}-y_{\omega_{[h+1]}}) \non
&& \cdot \sinc(y_{\omega_{[i-1]}}-y_{\omega_{[i+1]}}) \mid_{y_n=0} \dd \yv'_{n} \non
&=& \int_{\RR^{k-2}} \prod_{h=1}^{p-1} \sinc(y_{\omega'_h}-y_{\omega'_{[h+1]}})\mid_{y_n=0} \dd \yv'_n = v(\omegav')
\end{eqnarray*}
where $\yv'_n$ and $\omegav'$ have been obtained from $\yv_n$ and $\omegav$ by removing their $j$-th and $i$-th element, respectively.
Obviously $\yv'_n$ has size $k-1$ and $\omegav'$ has size $p-1$.
Let $\Pc_j$ be such that: $\Pc_j={i,[i+1]}$, i.e., $\omega_i=\omega_{[i+1]}=j$. Then,
\begin{eqnarray*}
&&\hspace{-6ex} v(\omegav) = \int_{\RR^{k-1}} \prod_{h=1}^p \sinc(y_{\omega_h}-y_{\omega_{[h+1]}})\mid_{y_n=0} \dd \yv_n \non
&=& \int_{\RR^{k-1}} \prod_{\substack{h=1 \\ h\neq i}}^p
  \sinc(y_{\omega_h}-y_{\omega_{[h+1]}}) \non
&&\qquad \cdot \sinc(y_{\omega_{i}}-y_{\omega_{[i+1]}}) \mid_{y_n=0}  \dd \yv_{n} \non
&=& \int_{\RR^{k-1}} \prod_{\substack{h=1 \\ h\neq i}}^p \sinc(y_{\omega_h}-y_{\omega_{[h+1]}})
    \sinc(y_j-y_j) \mid_{y_n=0}  \dd \yv_{n} \non
&=& \int_{\RR^{k-1}} \prod_{\substack{h=1 \\ h\neq i}}^p \sinc(y_{\omega_h}-y_{\omega_{[h+1]}})\mid_{y_n=0}  \dd \yv_{n} \non
&=& \int_{\RR^{k-1}} \prod_{h=1}^{p-1}\sinc(y_{\omega'_h}-y_{\omega'_{[h+1]}})\mid_{y_n=0}  \dd \yv_{n} =v(\omegav')
\end{eqnarray*}
where $\omegav'$ has been obtained from $\omegav$ by removing its $i$-th element.

\section{Proof of Lemma~\ref{th:mp}}
\label{app:mp}
In order to prove Lemma~\ref{th:mp}, we first note that $\Omega_{p,k}$ may 
contain both crossing and non-crossing 
partitions~\cite{DershowitzZaks}.

\paragraph{Non-crossing partitions}
Every non-crossing partition contains at least a singleton or a subset with adjacencies,
and therefore can be reduced by using the rules in Lemma~\ref{th:reduction_rules}.
After reduction, the resulting partition is still non-crossing, thus it can be further reduced till the empty set
is reached. It follows that the non-crossing partition $\omegav \in\Omega_{p,k}$ contributes to the expression of $\EE[\lambda_{p,k}^p]$ with
a coefficient $v(\omegav)=1$.

\paragraph{Crossing partitions}
Recall that, in general, the coefficient $v(\omegav)$ defined 
in~(\ref{eq:volume}) can be obtained
by counting the solutions of the system of equations:
$\Wm\ellv = \zerov$
where the $k(\omegav) \times p$ matrix $\Wm$ contains the coefficients of the $k(\omegav)$ constraints in~(\ref{eq:constraints}).

If $\omegav \in\Omega_{p,k}$ is a crossing partition, then
\begin{itemize}
\item $k(\omegav)\ge 2$ (by definition, a partition with $k(\omegav)= 1$ is always non-crossing)
\item it contains at least two subsets $\Pc_j$ and $\Pc_{j'}$, with 
$j\neq j'$, which are crossing.
\end{itemize}
Some crossing partitions can be reduced by applying the rules in Lemma~\ref{th:reduction_rules} but, 
even after reduction, they remain crossing.

Let us now focus on the crossing subset $\Pc_j$ of a partition
$\omegav$ which has been reduced by applying the rules in Lemma~\ref{th:reduction_rules}.
Without loss of generality, we assume that $|\Pc_j|=h$, i.e. the
partition $\Pc_j$ contains $h$ elements with $h\ge 2$ since $\Pc_j$ is not
a singleton.
Then, by definition of the matrix $\Wm$ (see
Appendix~\ref{app:constraints}) its $j$-th row, $\rv_j\Tran$, 
contains $h$ entries with value 1, $h$ entries with value $-1$ and $p-2h$ zeros.
We then build the $2\times p$ matrix $\widetilde{\Wm}$ as
$\widetilde{\Wm} = [\rv_j, -\rv_j]\Tran$.
Notice that $\widetilde{\Wm}$ has rank 1 and the system of equations $\widetilde{\Wm}\ellv=\zerov$
contains the constraints induced by a partition $\widetilde{\omegav}=[1,2,\ldots, 1,2]$ with $2h$ entries.
Since the system of equations $\widetilde{\Wm}\ellv=\zerov$ 
contains a reduced set of constraints with respect to $\Wm\ellv=\zerov$
and, thus, a larger number of solutions, it follows that 
$v(\omegav) \le v(\widetilde{\omegav})$.

It is straightforward to show that for a partition such as
$\widetilde{\omegav}$, with $k(\widetilde{\omegav})=2$, the
coefficient $v(\widetilde{\omegav})$ is given by Lemma~\ref{th:volume}
as $v(\widetilde{\omegav}) = \int_{\RR} \sinc(y)^{2h} \dd y$. This is
a decreasing function of $h$ and since $h\ge 2$ we have: 
\[ v(\widetilde{\omegav}) = \int_{\RR} \sinc(y)^{2h}\dd y \le
\int_{\RR} \sinc(y)^{4}\dd y = \frac{2}{3} \]
Therefore, we conclude that $v(\omegav) \le v(\widetilde{\omegav}) < 1$.

\paragraph{Crossing and non-crossing partitions}

Let $\Omega_{p,k}^{c}, \Omega_{p,k}^{n}\subset \Omega_{p,k}$ 
be, respectively, the set of crossing and non-crossing
partitions of $\Omega_{p,k}$, with $\Omega_{p,k}^{c} \cap \Omega_{p,k}^{n} = 
\emptyset$ and $\Omega_{p,k}^{c} \cup \Omega_{p,k}^{n} = \Omega_{p,k}$.
Then,
\begin{eqnarray*}
&&\hspace{-6ex}\EE[\lambda_{\infty,\beta}^p]
= \lim_{d \rightarrow +\infty}\sum_{k=1}^p  \beta^{p-k}\sum_{\omegav \in \Omega_{p,k}} v(\omegav)^d \non
&=& \lim_{d \rightarrow +\infty}\sum_{k=1}^p \beta^{p-k} \left(\sum_{\omegav \in \Omega_{p,k}^{c}} v(\omegav)^d + \sum_{\omegav \in 
\Omega_{p,k}^{n}} v(\omegav)^d \right) \non
&\overset{(a)}{=}& \sum_{k=1}^p \beta^{p-k} |\Omega_{p,k}^{n}| 
\end{eqnarray*}
where the equality $(a)$ is due to the fact that 
for non-crossing partitions $v(\omegav)=1$, while
for crossing partitions $v(\omegav)<1$ and, hence, 
$\lim_{d \rightarrow +\infty}v(\omegav)^d=0$.
In~\cite{Stanley}, it can be found that the number of non-crossing partitions of size 
$k$ in a $p$-element set is given by the Narayana 
numbers $T(p,k)= |\Omega_{p,k}^{n}|$
and therefore $\EE[\lambda_{\infty,\beta}^p] =  \sum_{k=1}^p T(p,k) \beta^{p-k}$
are the Narayana polynomials.
In~\cite{MarcenkoPastur}, it is shown that the Narayana polynomials
are the moments of the Mar\v{c}enko-Pastur distribution. 

\section{Proof of (\ref{eq:mseinf_LMMSE3})}
\label{app:proof11}
We show that when the LMMSE filter is used, the expression of the
asymptotic MSE is given by~(\ref{eq:mseinf_LMMSE3}).
Indeed, by using~(\ref{eq:MSEinf-persample}),~(\ref{eq:mse}),
(\ref{eq:hat_a}), and~(\ref{eq:B}) we have:
\[ \MSEinf =\limbeta\frac{1}{\sigma^2_a(2M+1)^d}\EE_\Xc\EE_{\av,
  \nv}\left[\|\Am_d^{-1}\Gm_d\pv-\av\|^2\right] \]
where $\Am_d = \Rm_d +\alpha \Id$ and
$\Rm_d = \Gm_d\Gm_d\Herm$. Substituting~(\ref{eq:p}) in the above
expression and assuming $\EE[\av\av\Herm]=\sigma^2_a\Id$ and $\EE[\nv\nv\Herm]=\sigma^2_n\Id$,
we get
\begin{eqnarray*}
&&\hspace{-6ex}\frac{1}{\sigma^2_a}\EE_{\av,\nv}\left[\|\Am_d^{-1}\Gm_d\pv-\av\|^2\right] \non
&=& \trace\left\{(\Am_d^{-1}\Rm_d-\Id)(\Am^{-1}\Rm_d-\Id)\Herm +
\alpha\Am_d^{-1}\Rm_d\Am_d^{-1}\right\} \non
&=& \trace\left\{\alpha(\Rm_d +\alpha \Id)^{-1}\right\} =
\trace\left\{\alpha\beta(\Tm_d +\alpha\beta \Id)^{-1}\right\}
\end{eqnarray*}
where $\Tm_d = \beta\Rm_d$. Let us consider an analytic function
$g(\cdot)$ in $\RR^+$. Let $\Xm =  \Um\Lambdam\Um\Herm$ be a random
positive definite Hermitian $n \times n$ matrix, where $\Um$ is the
eigenvectors matrix of $\Xm$ and $\Lambdam$ is a diagonal matrix
containing the eigenvalues of $\Xm$. By using the result for
symmetric matrices in~\cite[Ch. 6]{Bellman} combined with the result
in~\cite[pag. 481]{NumericalRecipes}, we have: 
$\lim_{n\rightarrow \infty}\frac{1}{n}\trace\EE [g(\Xm)] =
\EE_{\lambda}[ g(\lambda)]$ where the random variable $\lambda$ is
distributed as the asymptotic eigenvalues of $\Xm$. It follows that
\[ \limbeta\frac{\trace\EE_\Xc\left[\alpha\beta(\Tm_d
  +\alpha\beta \Id)^{-1}\right]}{(2M+1)^d} = \EE_{\lambda_{\beta,d}}\left[\frac{\alpha\beta}{\lambda_{\beta,d}
    +\alpha\beta} \right] \]


\end{document}